# Online Scheduled Execution of Quantum Circuits Protected by Surface Codes

Alexandru Paler
Institute for Integrated Circuits, Johannes Kepler University Linz
Linz Institute of Technology
Altenberger Straße 69, 4040 Linz, Austria

Austin G. Fowler
Google Inc.
Santa Barbara, California 93117, USA

Robert Wille
Institute for Integrated Circuits, Johannes Kepler University Linz
Altenberger Straße 69, 4040 Linz, Austria

**Abstract**

Quantum circuits are the preferred formalism for expressing quantum information processing tasks. Quantum circuit design automation methods mostly use a waterfall approach and consider that high level circuit descriptions are hardware agnostic. This assumption has lead to a static circuit perspective: the number of quantum bits and quantum gates is determined before circuit execution and everything is considered reliable with zero probability of failure. Many different schemes for achieving reliable fault-tolerant quantum computation exist, with different schemes suitable for different architectures. A number of large experimental groups are developing architectures well suited to being protected by surface quantum error correcting codes. Such circuits could include unreliable logical elements, such as state distillation, whose failure can be determined only after their actual execution. Therefore, practical logical circuits, as envisaged by many groups, are likely to have a dynamic structure. This requires an online scheduling of their execution: one knows for sure what needs to be executed only after previous elements have finished executing. This work shows that scheduling shares similarities with place and route methods. The work also introduces the first online schedulers of quantum circuits protected by surface codes. The work also highlights scheduling efficiency by comparing the new methods with state of the art static scheduling of surface code protected fault-tolerant circuits.

# 1 Introduction

Optimisation of fault-tolerant quantum circuits is not thoroughly investigated, because it is still difficult to agree on a common denominator technology for implementing such circuits. It is agreed that state of the art quantum computing architectures have to mitigate the unreliability of the quantum hardware by executing fault-tolerant quantum circuits. At the same time, consensus exists about the fact that assembling fault-tolerant quantum circuits comes at the cost of a high resource overhead dependant on the hardware (physical) unreliability and computational (logical) reliability.

A fault-tolerant quantum circuit operates at a logical layer and is obtained from an arbitrary circuit (operating at a physical layer) by using techniques based on quantum error correcting codes (QECC). The logical circuit layer abstracts physical layer resources: logical qubits are abstractions of physical qubit sets, and logical quantum gates abstract (sub)circuits of physical gates.

*Resource optimality* is concerned, in the context of this work, with the overheads introduced by fault-tolerance: space and time. Space overhead expresses the additionally required physical resources (e.g. hardware), and time overhead is dictated by the complexity of the fault-tolerance mechanisms (e.g. how long it takes to implement them). Furthermore, assuming that the quantum circuit formalism is well-known, it is possible to classify quantum circuit elements into operations (gates, subcircuits) and wires (qubits). As a result, as shown in Fig. 1, one can abstract each operation as a box. The circuit diagram can be further transformed into a *box diagram*, similar to the one from Fig. 2, by parameterising operation box dimensions with values representing the associated space and time overheads.

At this point, fault-tolerant quantum circuit optimisation can be defined as the task of *reducing overheads without impacting circuit fault-tolerance.*

Considering that hardware failure rates cannot be easily lowered (technological difficulties), there are at least two optimisation strategies: 1) either more capable QECCs are used (information theoretic perspective), or 2) the circuit operations are (re)arranged in a manner which utilises less resources (circuit design automation perspective). This work focuses on the latter approach, and argues that an efficient optimisation can be performed only at circuit run time because, as to be shown in the following, some circuit operations have a probabilistic nature: did the operation fail or not?; does the operation need to be repeated or not? Even for deterministic circuit operations, the quantum computing hardware needs to be dynamically allocated.

This work recognises that the optimisation of some fault-tolerant quantum circuits, as presented in the *Chosen Approach* section, is a scheduling problem related to place and route methods. It is shown that online scheduling is a necessity for resource efficiency when surface QECCs [6] are used.

The work includes two online scheduling algorithms, and their purpose is to show the optimisation constraints existing in a general framework. At the same time, the algorithms are used to motivate the study of fault-tolerant quantum circuit optimisation.

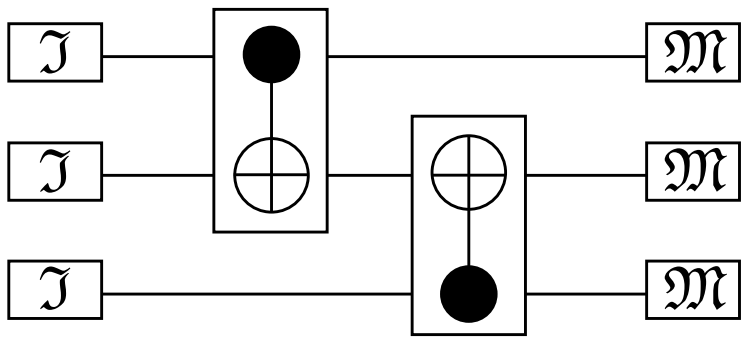

Figure 1: A quantum circuit where eight operations abstracted as boxes (three initialisations, two CNOT gates and three measurements) are applied to three qubits (parallel wires).

# 2 Background

The discussion is initiated by showing that the quantum circuit formalism has an intrinsic static structure which does not capture one of the key characteristics of fault-tolerant circuits: their potential probabilistic nature (see *Alternative Approaches* section).

## 2.1 Quantum Circuit Diagrams

Quantum circuits are the preferred formalism for expressing quantum information processing. The circuits, similarly to their classical counterparts, have inputs and outputs but, in contrast, consist of quantum gates operating on quantum bits (qubits). Additionally quantum circuits have their particularities: the number of inputs and outputs is always equal, and the circuits do not include any FANIN or FANOUT operations (quantum information cannot be copied). For these reasons, a quantum circuit is abstracted as a set of parallel wires (representing qubits) interrupted by gates, where information processing is executed from left to right. The inputs are on the left side of the circuit, and the outputs on the right side.

An imagined time axis parallel to the wires can be associated to each quantum circuit, and the position of each operation is an indication of its ordering within the circuit. Therefore, wires are diagrammatic representations of qubit time lines. There are three operation types: initialisation, gates and measurements. The quantum circuit terminology refers to qubit initialisations when setting input values (qubit states), and qubit measurements when reading computed results at the outputs.

As illustrated in Fig. 1, all circuit operations can be abstracted through *boxes*. Initialisations have only output wires, measurements only input wires, and gates have an equal number of input and output wires.

Without loss of generality, this work abstracts all the details about the dimensionality of a quantum circuit's state space, the possibilities and effects of initialisations and measurements, and the specific quantum phenomena implemented. It suffices to focus on circuit elements and how their fault-tolerant implementation is used by a quantum computer.

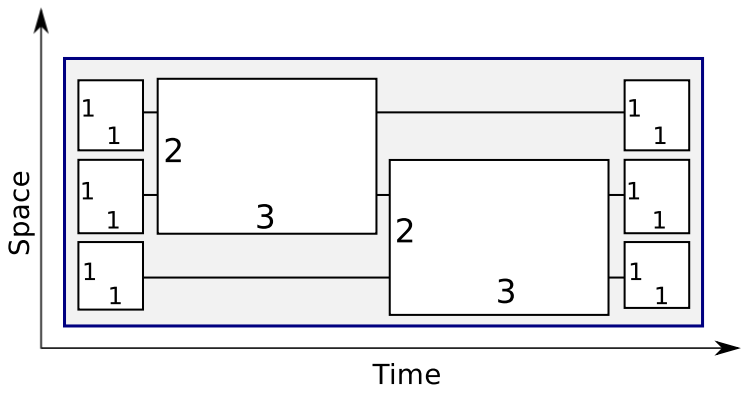

Figure 2: Operation costs: Each operation type from Fig. 1 has a space and time cost. For example, the CNOT has space cost 2 and time cost 3. The area of the blue bounding box expresses the required circuit resources (in this case $8 \times 3 = 24$). The offered costs have only an illustrative purpose, and realistic ones are computed by choosing specific QECCs.

## 2.2 Chosen Approach: Probabilistic and ICM Fault-Tolerant Circuits

An analysis of fault-tolerant circuits requires more than translating a circuit diagram into a box diagram, and the following analysis uses ICM, a general form of fault-tolerant circuits [14]. Such circuits include only single qubit (i)nitialisations, the (C)NOT gate (the two qubit gate from Fig. 1) and single qubit (m)easurements. ICM circuit universality is achieved through two types of initialisations, denoted in the following as *basis initialisations* and *injected initialisations*. All three operation types are addressed when assembling fault-tolerant quantum circuits.

Fault-tolerance is achieved, in general, by mitigating errors, which are the manifestation of hardware failures taking place during the execution of any circuit element. Logical circuit elements are constructed in a manner such that a circuit is fault-tolerant against $t$ errors if failures in $t$ elements result in at most $t$ errors per logical qubit [11]. Therefore, as long as hardware failure rates are below a certain threshold [11], QECCs can be used to increase computational reliability.

In particular, ICM circuits can be protected with the surface QECC [6], which is one of the alternatives to cope with the high failure rates of current quantum hardware [3]. Most ICM circuit operations can be executed on quantum hardware with failure rates below the necessary failure threshold, but the main issue is that injected initialisations [5] require distillation procedures which probabilistically fail [6]. However, distillations are *heralded*, meaning that it is known if they failed or not. Therefore, these initialisations and their distillations have to be repeated until a successful one has been executed.

The methods presented in this work are specifically targeted at surface code fault-tolerant quantum computing architectures.

### 2.3 Alternative Approaches: Non-Probabilistic

The herein discussed online schedulers are particularly concerned with the probabilistic behaviour introduced by the surface QECCs. Such codes require the distillation of injected initialisations. This work makes it possible to use topological cluster states (adequate for optics, optical lattices, and other architectures where qubits can be lost) and the surface code (adequate for architectures where qubits cannot be lost). Nevertheless, surface QECCs are not the only technical possibility for protecting fault-tolerant quantum circuits.

One option is to use QECCs which have the advantage that they do not require probabilistic distillation procedures of injected initialisations, but have the disadvantage that they introduce difficult to achieve hardware requirements. For example, long range interactions are required by [1, 4, 12], while [1, 8] introduce large hardware overheads which are, at least for the first generations of quantum architectures, prohibitive.

There exists also an intermediary technical path, where particular QECCs eliminate the requirement of certain distillations (e.g. [6]). As a result, some non-Clifford gates are deterministic (do not require distillation), while others are not (require distillation). Such QECC families of codes are of topological nature (not surface QECC), but would still require some online scheduling.

The alternative approaches contrast to the conservative approach of this work. The practicality of non-surface QECCs has not been studied to show how to overcome their difficulties. Distillation procedures could be theoretically circumvented, but there is ongoing research on how to apply this into practice. The alternative approaches are very promising, but not thoroughly investigated as the surface QECCs and, therefore, the methods presented in this work are tailored to the specifics of the latter.

## 3 Problem Statement

A quantum computer has $m$ available qubits, and $n$ operations forming an ICM fault-tolerant quantum circuit need to be executed on the computer. The space and time costs of each ICM operation are known. The circuit includes $n_i$ injected initialisations with an individual probability of distillation failure $p_f$. The entire computation expressed by the circuit is allowed to fail with a probability of at most $p_c$ where $p_f > p_c$.

Schedule all $n$ operations so that optimal space (qubits) and time resources are required.

### 3.1 Optimisation Objective Function

The objective function to minimise is related to a *bounding box* where height is space (qubits) and width is time. For example, Fig. 3, 4 and 5 illustrate the possible bounding boxes after scheduling the circuit from Fig. 1 having probabilistic distillations. Consequently, each circuit operation has a bounding box determined by space and time costs (e.g. Fig. 2), and a scheduled circuit has its own bounding box which is the convex hull of all operation bounding boxes.

### 3.2 Synthesis, Scheduling, Place and Route: Analogies

Synthesis and scheduling share certain conceptual analogies, discussed in the following.

The scheduling terminology considers jobs to be processed by machines under certain constraints. In the case of fault-tolerant quantum circuits, machines are qubits and jobs are operations. Each operation to be scheduled is a single qubit initialisation, a CNOT or a single qubit measurement. Multiple constraints can be analysed, but the central ones are spatial and precedence. Spatial constraints refer to the amount of available qubits (wires). Precedence constraints are related to the partial ordering between operations (time).

The most common objective function that is optimised during classical process scheduling is the *makespan*. This is the length of the schedule, or equivalently the time when the last job is completed [17]. Fault-tolerant quantum circuits have an intrinsic space cost (overhead) and the makespan is not sufficient for an objective function. Quantum circuit cost models were proposed for example in [16]. In general the models are technology dependent and related to a space time cost, but can be classified into: focusing more on space overheads (e.g. number of one qubit and CNOT gates, number of garbage qubits), and focusing more on time overheads (e.g. circuit depth).

As a conclusion, scheduling fault-tolerant quantum circuits is a variant of classical circuit place and route algorithms, as exemplified by Fig. 2. Therefore, the *bounding box* associated to a quantum circuit is the natural objective to minimise. Further details about space and time optimisation of quantum circuits can be found in e.g. [10, 18].

### 3.3 Qubit State Movement

Some conceptual differences seem to exist between a synthesis result (circuit diagram) and a scheduling result (box diagram), as seen for example in Fig. 1 and Fig. 2. However, the conceptual differences between synthesis and scheduling disappear if the latter is understood as *place and route*. Circuit synthesis is expected to take operations, associated beforehand to a set of qubits (wires), and to just place them on the qubits in a given order. This means, in terms of a scheduling algorithm, that jobs would have a strict machine preferences. This seems very limiting. In contrast, the target of scheduling is to assign operations to qubits, assuming that any qubit could be used as long as it is not used by another operation. The impression is that scheduling is not limiting, assumes that almost all jobs have no machine preference, and does not generate generate correct computations.

In the context of quantum circuits there is a major distinction to be made between a qubit's wire and a qubit's state: operations are affecting the state and not the wire. Wires are only diagrammatic abstractions of a relative operation ordering, and it is possible to move a state from one wire to another (e.g. state swapping or state teleportation [11]). More specifically, in a circuit diagram state movements could be implemented using SWAP subcircuits, but in a box diagram, state movements are indicated by just bending the connections (wires) between the operation boxes as seen for example in Fig. 3.

State movement is also related to the malleability [17] of an operation, which means if the operation can be scheduled on fewer qubits at the cost of increasing processing time. A single CNOT gate is clearly not malleable, because it is a parallel operation applied on two qubits. However it is valid to assume that certain operations in a large scale circuit are ICM subcircuits (e.g. a two qubit full adder) instead of gates, and such operations could be malleable.

This observation shows, once more, that online scheduling is very similar to a classical circuit place and route problem with bus (interconnections between operations) constraints [7].

### 3.4 Online Methods: Motivation

The notion of an online algorithm formalises the scenario where the algorithm does not have access to the whole input set [17]. Online scheduling is scheduling with incomplete information at certain points, and scheduling decisions are irreversible.

Online scheduling methods are required for at least two practical reasons: a technology specific one, and a more general one related to how QECCs are applied to circuits.

Throughout this work, for simplification purposes, the term initialisation will refer to the entire process consisting of injecting magic states and increasing their fidelity through distillation procedures.

The first motivation is that from a technology point of view, circuit synthesis, optimisation and execution of ICM circuits protected by surface QECCs is influenced by the probabilistic nature of injected state distillation. Therefore, for surface code based quantum computing architectures, resource overhead optimisation should be performed using online algorithms, because the complete input set of operations is not known beforehand: which initialisations need to be repeated? In other words, the current operation can be scheduled only after the preceding ones were successfully executed, because some of them are probabilistic (and heralded) and require repetitions. There are known methods to circumvent the necessity of the probabilistic distillations (see *Alternative Approaches* section), but they are not applicable to the way surface QECCs are used.

The more general motivation for achieving resource consumption optimality through online scheduling, is that it is reasonable to assume that in a large scale quantum computer the QECC strength (without restriction to a particular QECC) needs to be dynamically adapted to the fluctuating failure rate of the hardware. This implies that, in practice, the cost of each error corrected circuit operation is fluctuating, and quantum computations need to be dynamically adapted to fit into the available computational resources (space and time). This would be a very similar approach to how computational resources are shared between processes on classical computers by the operating system schedulers and resource managers.

The current paper will focus entirely on ICM circuits protected by the surface QECC, such that online methods need to solve the placement (scheduling) of probabilistic distillations.

# 4 Offline Scheduling Solution (ASAP)

An unoptimal solution to the scheduling problem was offered in [13], where the authors considered the synthesis of arbitrary quantum circuits protected by the surface QECC. Although not directly stated by [13], circuit synthesis and scheduling are tightly related. Synthesis introduces a relative point of time for the execution of an operation (the operation precedence), while scheduling chooses an exact point of time for the execution. Synthesis performs the following operations: a) introduces ancilla qubits (if required); 2) associates operations (gates from a specific gate set) to qubits (wires). Each gate placement determines an operation precedence: execute a gate no sooner than the preceding gate placed on the same wire, and no later than the succeeding gate. Precedence is generally formulated using directed acyclic graphs on the operations; each directed edge indicates that one operation has to be scheduled before another one [17, 9]. In this work, directed edges are the wires interconnecting the operations.

The work of [13] recognised that initialisations are probabilistic and heralded, but it did not recognise that the therein presented ICM circuit synthesis was an *as soon as possible offline scheduling* (ASAP). Their scheduling (synthesis) is based on a model using explicitly parameters equivalent to $n_i$, $p_f$ and $p_c$ from the problem statement: a worst case amount of additional initialisations was synthesised (scheduled) into the circuit, so that at least $n_i$ were expected to succeed in order to achieve $p_c$. Successful initialised states were used in the circuit (connected to the circuit).

The unoptimality of the ASAP approach of [13], sketched in Fig. 3, originates from the fact that it executes all distillations in parallel (does not consider any space constraints). This is not space efficient, because, not all initialisations are required right at the beginning of the circuit, and could be executed just before the operation which succeeds them.

Furthermore, ASAP is an offline scheduling algorithm, because all operations have a strict ordering along the time axis as seen in Figure 3: 1) initialisations are placed at time coordinate zero, 2) all other non-probabilistic operations have a time coordinate strictly determined only by their precedence. In contrast, time axis coordinates are not strict if initialisations are executed just before the operations, and this steers the discussion of this work towards online scheduling methods.

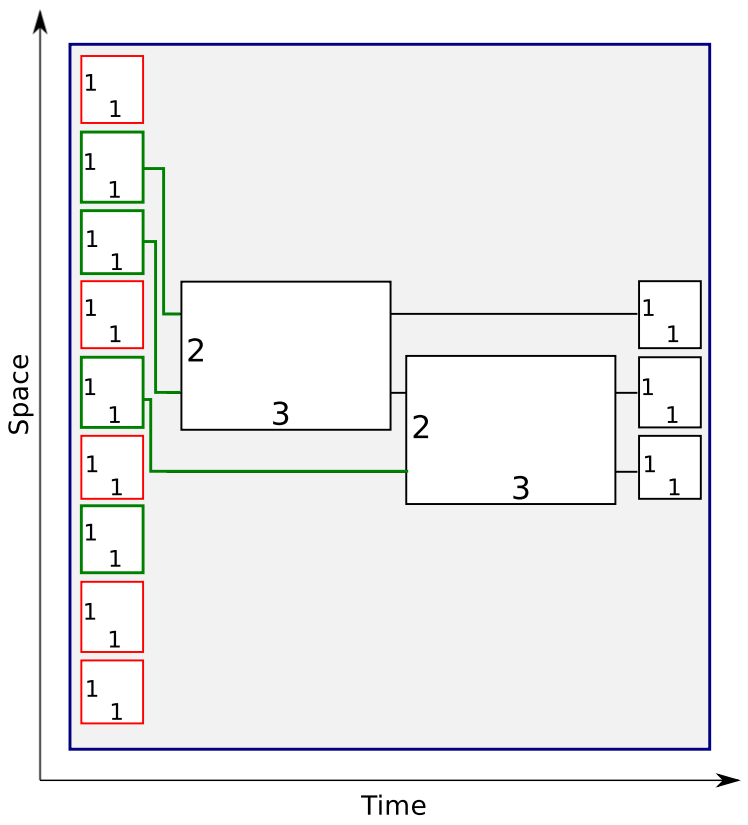

Figure 3: ASAP scheduling of ICM circuits. Multiple probabilistic distillations are executed before any other circuit operation is performed. Successful distillations (green) are used by the circuit, while failed ones (red) are discarded. For this example the total resources required are $8 \times 9 = 72$.

# 5 Online Scheduling Algorithms

The following two online scheduling algorithms of fault-tolerant quantum circuits are discussed using the work from [7] as a foundation. The algorithms are formulated for the extreme optimisation cases: either only space, or only time.

A hybrid approach (combining both space and time optimisation) is possible, but given the current state of quantum technologies (reduced number of qubits), the only practical option is an aggressive online optimisation of space (reduce number of circuit qubits as much as possible, without considering too much the time penalty introduced). Accordingly, only the extreme cases will be discussed in the following.

## 5.1 Time Constrained ALAP (ALAPT)

The first proposed online algorithm is a time constrained online scheduling that places distillations in parallel (Fig. 4). This starts from the (for current technologies, unrealistic) assumption that there is enough hardware available (ideally $m = \infty$) for a very large number of initialisations.

The algorithm proceeds as follows: each time an initialisations is required one is placed on the specified qubit, and additional ones on ancilla qubits, so that all initialisations will be executed in parallel. Successful distillations are indicated by their heralded execution result, and in the worst case it is guaranteed that one succeeds. The successful one will be used (if necessary by qubit state movement) on the specified logical qubit.

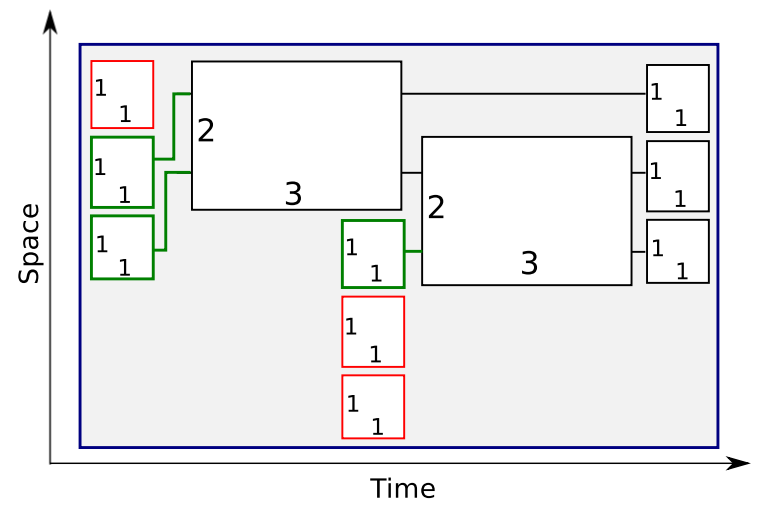

Figure 4: Time constrained ALAP scheduling of ICM probabilistic distillations. Multiple distillations are executed in parallel right before the circuit requires at least one. Successful distillations (green) are used, while failed ones (red) are discarded. The state of unused successful initialisations is used at a later point of time, thus reducing space resource consumption. For this example the total resources required are $8 \times 5 = 40$.

The online property stems from the following fact: it may happen that more than one distillation succeeds. Thus, successive operations requiring an initialised qubit can check first if previous successful initialisations were executed but not used, and otherwise schedule and execute another round of parallel ones. The advantage of the former option is that it saves the time required for parallel executions.

## 5.2 Space Constrained ALAP (ALAPS)

The second online scheduling algorithm assumes that space is constrained, but time is infinite (Fig. 5). This assumption is the most realistic for state of the art quantum computing architectures. A circuit will be scheduled and executed as long as it has at most $m$ qubits (the limited hardware support). Under this assumption, it is possible to devise two ALAP strategies: a) repeat until an initialisation does not fail; 2) schedule a sequence of additional initialisations such that $p_c$ (overall computational failure probability, see problem statement) can be guaranteed.

The first strategy is straightforward and guaranteed to lead to the $p_c$ computational reliability. The second strategy may generate, similarly to ALAPT, multiple successful initialisations in a sequence, and successive operations are allowed to check first if previous successful initialisations were executed but not used. In the worst case both strategies require the same number of sequential initialisation trials.

## 5.3 Practical Implementation

Online scheduling has to be understood as a synthesis, optimisation and execution method coupled to the feedback of the quantum hardware: each part of a fault-tolerant circuit is synthesised only after previous parts were successfully executed. Synthesis, optimisation are not a sequential process any more, similar to a waterfall procedure, but more dynamic one which includes feedback: partial synthesis is followed by partial optimisation in a loop until the entire circuit is synthesised, optimised and executed. The probabilistic nature of surface code protected fault-tolerant circuits forces automated design methods to not focus on global circuit optimums, but on achieving sufficiently good local optimums. Compared to classical synthesis and optimisation (without hardware feedback), online scheduling introduces no relevant computational overhead, but its true cost is mirrored in local (instead of global) perspective it has on the circuit.

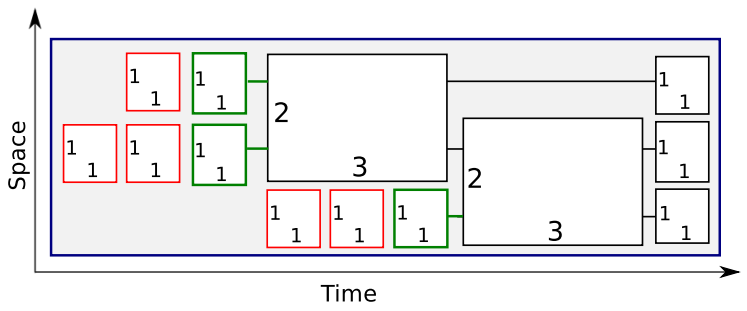

Figure 5: Space constrained ALAP scheduling of ICM probabilistic initialisations. Multiple initialisations are executed sequentially until one succeeds (green). For this example the total resources required are $10 \times 3 = 30$. Although less resources are used compared to the time constrained example, the execution of this scenario takes longer to execute $(10 > 8)$.

# 6 Discussion

Implementations of ALAPT and ALAPS online scheduling are evaluated against the offline ASAP scheduling. The benchmark is formed of circuits, from the RevLib library, consisting entirely of multi-controlled Toffoli [11] gates (MCT). The schedulers were implemented as extensions of the software presented in [13]. The evaluation results in Table 1 illustrate the worst case bounding boxes obtained when scheduling ICM circuits generated from the corresponding RevLib MCT circuits.

## 6.1 Evaluation Setup

MCT circuits are not fault-tolerant, and their fault-tolerant ICM form was obtained by decomposing n-controlled ($n \geq 3$) Toffoli gates into ancilla qubits and 2-controlled Toffoli gates [2], further expressed into ICM using [13].

The resulting ICM circuits include two types of probabilistic initialisations: A-type and Y-type [5]. The columns A and Y in Table 1 indicate the corresponding numbers in each ICM circuit. The used Toffoli gate decomposition results in Y initialisations being twice as many than A initialisations (cf. columns A and Y). According to [6]: A-type distillation has time cost seven and wire cost 15, and Y-type initialisation has gate cost six and wire cost seven. All costs were increased by two, in order to permit qubit state movement (connect successfully distilled states to the circuit operations). Each CNOT gate was modelled with gate cost one and wire cost two.

Table 1: Scheduling Results

| | | | Optimised Synthesis | | | | | | | | | Unoptimised Synthesis | | | | | | | | |
|---|---|---|---|---|---|---|---|---|---|---|---|---|---|---|---|---|---|---|---|---|
| | | | ASAP | | | ALAPT | | | ALAPS | | | ASAP | | | ALAPT | | | ALAPS | | |
| Circuit | A | Y | T | S | BB | T | S | BB | T | S | BB | T | S | BB | T | S | BB | T | S | BB |
| 3_17_13 | 14 | 28 | 337 | 856 | 288472 | 650 | 94 | 61100 | 2018 | 26 | 52468 | 234 | 856 | 200304 | 442 | 172 | 76024 | 1759 | 104 | 182936 |
| 3_17_14 | 14 | 28 | 354 | 856 | 303024 | 684 | 94 | 64296 | 2052 | 26 | 53352 | 247 | 856 | 211432 | 561 | 172 | 96492 | 1802 | 104 | 187408 |
| 4_49_16 | 77 | 154 | 1695 | 3865 | 6551175 | 3491 | 100 | 349100 | 10935 | 32 | 349920 | 1185 | 3865 | 4580025 | 2744 | 552 | 1514688 | 10086 | 484 | 4881624 |
| 4_49_17 | 35 | 70 | 857 | 1888 | 1618016 | 1650 | 97 | 160050 | 5069 | 29 | 147001 | 594 | 1888 | 1121472 | 1231 | 299 | 368069 | 4445 | 231 | 1026795 |
| 4gt10-v1_81 | 49 | 98 | 1126 | 2553 | 2874678 | 2264 | 101 | 228664 | 6711 | 33 | 221463 | 760 | 2553 | 1940280 | 1535 | 385 | 590975 | 6230 | 317 | 1974910 |
| 4gt11_82 | 7 | 14 | 197 | 489 | 96333 | 357 | 94 | 33558 | 1025 | 26 | 26650 | 138 | 489 | 67482 | 280 | 132 | 36960 | 878 | 64 | 56192 |
| 4gt11_83 | 7 | 14 | 190 | 489 | 92910 | 338 | 92 | 31096 | 1002 | 24 | 24048 | 133 | 489 | 65037 | 275 | 132 | 36300 | 878 | 64 | 56192 |
| 4gt11_84 | 7 | 14 | 256 | 489 | 125184 | 408 | 92 | 37536 | 1076 | 24 | 25824 | 126 | 489 | 61614 | 266 | 132 | 35112 | 881 | 64 | 56384 |
| 4gt11-v1_85 | 7 | 14 | 256 | 489 | 125184 | 410 | 92 | 37720 | 1078 | 24 | 25872 | 126 | 489 | 61614 | 315 | 132 | 41580 | 983 | 64 | 62912 |
| 4gt12-v0_86 | 63 | 126 | 1241 | 3226 | 4003466 | 2723 | 106 | 288638 | 7667 | 38 | 291346 | 870 | 3226 | 2806620 | 2247 | 470 | 1056090 | 8320 | 402 | 3344640 |
| 4gt12-v0_87 | 63 | 126 | 1235 | 3226 | 3984110 | 2716 | 106 | 287896 | 7660 | 38 | 291080 | 862 | 3226 | 2780812 | 2239 | 470 | 1052330 | 8312 | 402 | 3341424 |
| 4gt12-v0_88 | 49 | 98 | 1061 | 2553 | 2708733 | 2206 | 106 | 233836 | 6784 | 38 | 257792 | 745 | 2553 | 1901985 | 1519 | 386 | 586334 | 6235 | 318 | 1982730 |
| 4gt12-v1_89 | 56 | 112 | 1283 | 2885 | 3701455 | 2584 | 104 | 268736 | 7487 | 36 | 269532 | 841 | 2885 | 2426285 | 1714 | 428 | 733592 | 7131 | 360 | 2567160 |
| 4gt13_90 | 35 | 70 | 735 | 1888 | 1387680 | 1551 | 101 | 156651 | 4352 | 33 | 143616 | 521 | 1888 | 983648 | 1210 | 301 | 364210 | 4480 | 233 | 1043840 |
| 4gt13_91 | 35 | 70 | 729 | 1888 | 1376352 | 1546 | 100 | 154600 | 4596 | 32 | 147072 | 515 | 1888 | 972320 | 1204 | 301 | 362404 | 4474 | 233 | 1042442 |
| 4gt13_92 | 21 | 42 | 550 | 1206 | 663300 | 1021 | 97 | 99037 | 3012 | 29 | 87348 | 323 | 1206 | 389538 | 622 | 217 | 134974 | 2651 | 149 | 394999 |
| 4gt13-v1_93 | 21 | 42 | 544 | 1206 | 656064 | 1012 | 98 | 99176 | 3030 | 30 | 90900 | 328 | 1206 | 395568 | 626 | 217 | 135842 | 2646 | 149 | 394254 |
| 4gt4-v0_72 | 70 | 140 | 1372 | 3550 | 4870600 | 3052 | 106 | 323512 | 9396 | 38 | 357048 | 962 | 3550 | 3415100 | 2107 | 512 | 1078784 | 8923 | 444 | 3961812 |
| 4gt4-v0_73 | 112 | 224 | 2392 | 5492 | 13136864 | 4978 | 106 | 527668 | 14928 | 38 | 567264 | 1672 | 5492 | 9182624 | 4399 | 764 | 3360836 | 15363 | 696 | 10692648 |
| 4gt4-v0_78 | 56 | 112 | 1246 | 2885 | 3594710 | 2516 | 104 | 261664 | 7335 | 36 | 264060 | 854 | 2885 | 2463790 | 2035 | 428 | 870980 | 7408 | 360 | 2666880 |
| 4gt4-v0_79 | 56 | 112 | 1232 | 2885 | 3554320 | 2502 | 104 | 260208 | 7321 | 36 | 263556 | 844 | 2885 | 2434940 | 2025 | 428 | 866700 | 7398 | 360 | 2663280 |
| 4gt4-v0_80 | 42 | 84 | 934 | 2229 | 2081886 | 1912 | 104 | 198848 | 5817 | 36 | 209412 | 644 | 2229 | 1435476 | 1323 | 344 | 455112 | 5339 | 276 | 1473564 |
| 4gt4-v1_74 | 77 | 154 | 1698 | 3865 | 6562770 | 3498 | 106 | 370788 | 10285 | 38 | 390830 | 1177 | 3865 | 4549105 | 2727 | 554 | 1510758 | 10191 | 486 | 4952826 |
| 4gt5_75 | 28 | 56 | 641 | 1547 | 991627 | 1276 | 100 | 127600 | 3923 | 32 | 125536 | 450 | 1547 | 696150 | 822 | 259 | 212898 | 3542 | 191 | 676522 |
| 4gt5_76 | 28 | 56 | 683 | 1547 | 1056601 | 1307 | 98 | 128086 | 4025 | 30 | 120750 | 434 | 1547 | 671398 | 904 | 259 | 234136 | 3542 | 191 | 676522 |
| 4gt5_77 | 42 | 84 | 922 | 2229 | 2055138 | 1901 | 100 | 190100 | 5767 | 32 | 184544 | 638 | 2229 | 1422102 | 1218 | 343 | 417774 | 5339 | 275 | 1468225 |
| 4mod5-bdd_287 | 28 | 56 | 612 | 1547 | 946764 | 1251 | 99 | 123849 | 3968 | 31 | 123008 | 429 | 1547 | 663663 | 826 | 260 | 214760 | 3542 | 192 | 680064 |
| 4mod5-v0_18 | 28 | 56 | 670 | 1547 | 1036490 | 1298 | 100 | 129800 | 3883 | 32 | 124256 | 464 | 1547 | 717808 | 797 | 258 | 205626 | 3542 | 190 | 672980 |
| 4mod5-v0_19 | 14 | 28 | 341 | 856 | 291896 | 658 | 95 | 62510 | 2022 | 27 | 54594 | 237 | 856 | 202872 | 401 | 174 | 69774 | 1750 | 106 | 185500 |
| 4mod5-v0_20 | 7 | 14 | 185 | 489 | 90465 | 338 | 92 | 31096 | 1006 | 24 | 24144 | 122 | 489 | 59658 | 205 | 132 | 27060 | 854 | 64 | 54656 |
| 4mod5-v1_22 | 7 | 14 | 180 | 489 | 88020 | 331 | 94 | 31114 | 995 | 26 | 25870 | 129 | 489 | 63081 | 212 | 132 | 27984 | 858 | 64 | 54912 |
| 4mod5-v1_23 | 28 | 56 | 666 | 1547 | 1030302 | 1295 | 100 | 129500 | 3938 | 32 | 126016 | 456 | 1547 | 705432 | 797 | 258 | 205626 | 3550 | 190 | 674500 |
| 4mod5-v1_24 | 14 | 28 | 363 | 856 | 310728 | 687 | 93 | 63891 | 2055 | 25 | 51375 | 244 | 856 | 208864 | 474 | 174 | 82476 | 1755 | 106 | 186030 |
| 4mod7-v0_94 | 56 | 112 | 1118 | 2885 | 3225430 | 2454 | 101 | 247854 | 7226 | 33 | 238458 | 780 | 2885 | 2250300 | 1750 | 427 | 747250 | 7131 | 359 | 2560029 |
| 4mod7-v0_95 | 56 | 112 | 1217 | 2885 | 3511045 | 2525 | 101 | 255025 | 7712 | 33 | 254496 | 846 | 2885 | 2440710 | 1795 | 427 | 766465 | 7131 | 359 | 2560029 |
| 4mod7-v1_96 | 56 | 112 | 1217 | 2885 | 3511045 | 2525 | 101 | 255025 | 7712 | 33 | 254496 | 846 | 2885 | 2440710 | 1795 | 427 | 766465 | 7131 | 359 | 2560029 |

The synthesis software [13] supports an optimisation technique called wire recycling [15]. Optimised circuits will use fewer ancilla and have a higher depth (take longer to execute), although the circuits are functionally and structurally equivalent to the unoptimised ones. Both the optimised and the unoptimised ICM circuits will thus an equal number of A-type and Y-type initialisations. Accordingly, Table 1 includes the results for scheduling optimised and unoptimised ICM versions of the same MCT circuit.

The scheduler evaluation is designed to solve the problem statement enounced earlier: circuits should fail at most with probability $p_c = 0.001$ while each initialisation (A or Y) will fail with $p_f = 0.2$. Each ICM circuit has $n_i$ initialisations with $p_f$ distillation failure probability. Additional initialisations ($s$) are necessary for achieving $p_c$, resulting in $n_t = s + n_i$ initialisations to schedule. The parameter $s$ is determined such that $1 - F(s, n_t, p_f) < p_c$:

$$F(s, n_t, p_f) = \prod_{k=0}^{s} \binom{n_t}{k} p_f^k (1 - p_f)^{n_t - k} \tag{1}$$

On the one hand, in the case of offline ASAP $n_i$ refers to the A and Y columns in Table 1. For example, circuit `3_17_13` has 14 A initialisations and 28 Y initialisations, implying that computational failure $p_c = 0.001$ is guaranteed when 12 additional A and 18 additional Y initialisations are used. On the other, ALAPT and ALAPS are online, and the number of additional initialisations is determined for $n_t = s + 1$, meaning that for each A or Y initialisation four additional ones ($s = 4$) are required in the worst case. ALAPT will schedule five initialisations in parallel, while ALAPS schedules five initialisations sequentially (cf. Fig. 4 and Fig. 5).

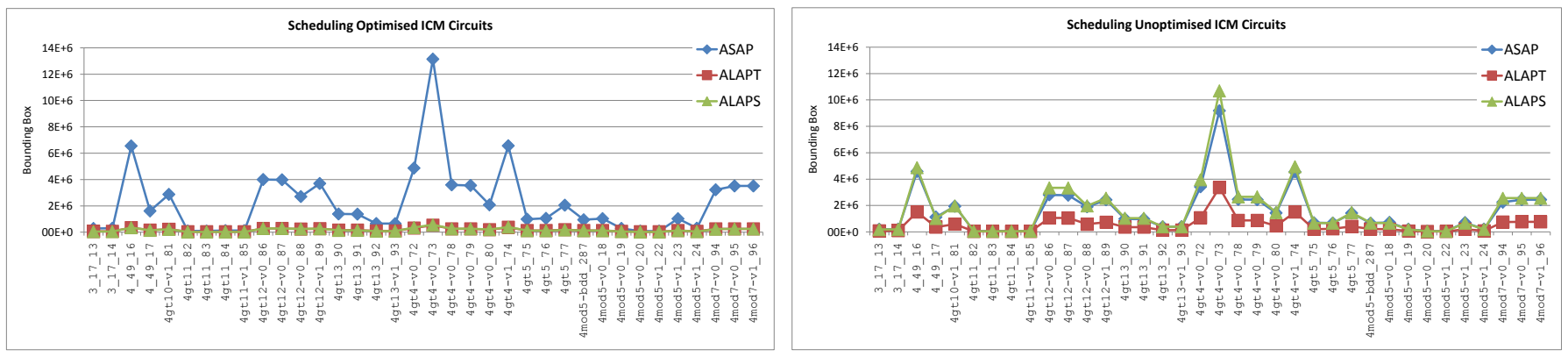

Figure 6: Scheduling optimised and unoptimised ICM circuits. Optimised circuits use less ancilla and the costs of initialisations dominate the overall amount of spatial resources required. For this reason ALAPT and ALAS, although performing much better than ASAP, generate bounding boxes of approximately the same area. In the case of unoptimised ICM circuits, ALAPS scheduling performs only slightly better than ASAP. This is because the spatial optimisation achieved by ALAPS is almost entirely counterbalanced by the introduced time penalty.

## 6.2 Simulation Results

The results in Table 1 indicate the time cost (column T), the space cost (column S) and the bounding box (column BB, time × space) for each ICM circuit instance (optimised and unoptimised synthesis) scheduled using ASAP, ALAPT and ALAPS. The results are interpreted from two perspectives. Optimised synthesis is the one using wire recycling.

### 6.2.1 Online Scheduling vs. Offline(ASAP)

The first conclusion is that online scheduling performs better than offline (ASAP) scheduling (cf. bounding boxes of the same circuit generated by ASAP and any online scheduling algorithm). This was to be expected by the very inefficient design of ASAP: all additional distillations were placed in parallel, the overall space requirement being dominated by these, and thus resulting in largely unoccupied bounding boxes (e.g. the empty space in Fig. 3). ASAP could have been improved by placing the additional distillations in parallel sequences, a layout similar to a matrix instead of a column vector. Nevertheless, any improvement of ASAP would reach only the average performance of an online scheduler: an online scheduled (synthesised) circuit requires new initialisations only if no previous successful ones exist (see description of ALAPT and ALAPS), but an offline generated version of the circuit does not have this option.

The number of ancilla qubits impacts ASAP scheduling performance in a negative sense. Optimised circuits use less ancillae at the cost of more execution time (cf. column optimised-ASAP-BB with column unoptimised-ASAP-BB). For example, circuit `3_17_13`, in its optimised version has a time cost of 337, and the unoptimised one only 234. However, using ASAP, the high number of additional initialisations dominates the space cost (cf. column optimised-ASAP-S with column unoptimised-ASAP-S are equal for all circuits), such that time overhead will influence bounding box size.

Online scheduling of optimised circuits delivers the best results when the circuit includes a high number of A and Y initialisations. For example, for the `4gt10-v1_81` circuit, the ASAP bounding box is more than 10 times larger than the ALAPT bounding box ($2874678/228664 \approx 12.6$, whereas for `3_17_13` the ASAP bounding box is only approximately five times larger than the ALAPT ($288472/61100 \approx 4.7$). The high number of A and Y initialisations is an indication that the circuit operates on a large number of ancillae even after optimisation, such that parallel execution of initialisations does not dominate the overall space costs.

Online scheduling of unoptimised circuits delivers the best results for ALAPT, while ALAPS results in bounding boxes marginally smaller than ASAP (Fig. 6). ALAPS uses almost eight times less space resources compared to ASAP (cf. columns unoptimised-ASAP-S and unoptimised-ALAPS-S), but introduces approximately the same overhead factor for time costs (cf. columns unoptimised-ASAP-T and unoptimised-ALAPS-T). For the same `3_17_13` circuit, ASAP bounding box is $238 \times 856$ and ALAPS bounding box is $1759 \times 104$. The ALAPS spatial optimisation is almost entirely counterbalanced by the introduced time penalty.

### 6.2.2 ALAPT vs. ALAPS

It was expected for ALAPT to result in shorter executions compared to ALAPS, and this is true for both optimised and unoptimised circuits. Similarly, due to its algorithmic design, ALAPS uses less space than ALAPT, but for optimised circuits both methods generate bounding boxes of approximately equal sizes: the time optimisation achieved by ALAPT is counterbalanced by the space required for ALAPS parallel initialisations; time is traded for space.

Online schedulers could be refined to further optimise ICM circuits, but the major costs are dictated by the distillation procedures and their probabilistic nature. The higher the probability of failure ($p_f$), the more parallel (ALAPT) or sequential (ALAPS) initialisations are required. The high cost of each initialisation is multiplied by the number of additionally required ones. Therefore, initialisations dominate the overall bounding box of an online scheduled optimised ICM circuit.

# 7 Conclusion

This work is the first to consider the optimisation of fault-tolerant quantum circuits as an online scheduling problem. Such circuits have been shown to be formed from only three types of operations (ICM: initialisations, CNOT gates and measurements). Large-scale fault-tolerant quantum computations can be achieved, in architectures based on surface QECCs, through efficient online scheduling methods. Two online scheduling algorithms were proposed and their potential was highlighted by comparing them against the offline state of the art ASAP scheduling existing in the literature.

Future work will focus on devising hybrid online scheduling algorithms which adaptively use space and time constraints. Furthermore, more refined metrics (complementing the bounding box concept) will be investigated. Due to the architecture of the most promising quantum computing architectures it is also envisioned to extend the presented scheduling methods to three dimensional spaces instead of two dimensional ones.

# 8 Acknowledgement

This work was supported by the Linz Institute of Technology project CHARON.